\documentclass[%
reprint,
superscriptaddress,
nofootinbib,
nobibnotes,
amsmath,amssymb,
aps, prd
]{revtex4-2}

\usepackage{subcaption}
\usepackage{graphicx}
\usepackage{dcolumn}
\usepackage{bm}
\usepackage{hyperref}
\usepackage{amsmath} 
\usepackage{physics}
\usepackage{amssymb}
\usepackage{soul}
\usepackage{aas_macros}
\usepackage{tabularx}
\usepackage{float}

\newcommand{\fDu}[1]{\stackon[-0.3ex]{$D^{#1}$}{\kern-1.0ex\scalebox{0.7}{$\circ$}}}

\newcommand{\zD}{{\raise1.0ex\hbox{${}^{\ \circ}$}}\!\!\!\!\!D}

\usepackage{orcidlink}

\graphicspath{{./}{figures/}{figures/res_1e6_pngs/}{figures/res_1e4_pngs/}{figures/res_1e3_pngs/}{figures/res_1e2_pngs/}}

\begin{document}

\preprint{APS/123-QED}

\title{Magnetic effects on fundamental modes in rotating neutron stars with a purely toroidal magnetic field}

\author{Anson Ka Long \surname{Yip}~\orcidlink{0009-0008-8501-3535}}
\email{kalongyip@cuhk.edu.hk}
\affiliation{Department of Physics, The Chinese University of Hong Kong, Shatin, N.T., Hong Kong}


\author{Tjonnie Guang Feng \surname{Li}~\orcidlink{0000-0003-4297-7365}}
\affiliation{Institute for Theoretical Physics, KU Leuven, Celestijnenlaan 200D, B-3001 Leuven, Belgium}
\affiliation{Department of Electrical Engineering (ESAT), KU Leuven, Kasteelpark Arenberg 10, B-3001 Leuven, Belgium }

\date{\today}

\begin{abstract}
Electromagnetic and gravitational-wave signals from neutron stars are shaped by rapid rotation and strong magnetic fields. 
Determining these properties is essential to interpret such signals, but current measurements are limited: rotation estimates rely on electromagnetic detections and assume uniform rotation, while inferring interior magnetic fields remains ambiguous due to a lack of direct observations. 
Measuring the excited fundamental modes of neutron stars in gravitational-wave signals offers a promising solution, as these modes encode information about stellar composition, structure, and dynamics.
Previous studies have examined the individual effects of rotation and magnetic fields on these modes, identifying magnetic suppression and establishing linear relations for the frequencies of the fundamental $l=0$ quasi-radial mode $f_F$ and $l=2$ quadrupolar mode $f_{^2f}$. 
However, few have investigated the combined influence of rotation and magnetic fields. 
Here, for the first time, we consider both rotation and a toroidal magnetic field to construct linear relations for quantifying $f_F$ and $f_{^2f}$, showing that their combined effects can be constrained by detecting these modes.
Using 2D axisymmetric simulations, we demonstrate that quasi-linear relations between $f_F$, $f_{^2f}$, stellar compactness $M/R$, and kinetic-to-binding energy ratio $T/|W|$ persist even with a toroidal magnetic field. 
The slope of these relations depends on the toroidal magnetization constant $K_\mathrm{m}$. Additionally, measuring the frequency ratio $f_{^2f}/f_F$ enables inference of $T/|W|$ and the maximum magnetic field strength $\mathcal{B}_\mathrm{max}$. 
Lastly, we show that differential rotation causes only minor deviations from predictions for uniform rotation. 
Thus, this work demonstrates that rotational and magnetic properties of neutron stars can be inferred from their fundamental modes.

\end{abstract}

\maketitle


\section{Introduction} \label{sec:intro}
Neutron stars exhibit a wide range of rotation periods $P$, spanning from a few ms to several s. 
For example, classical pulsars generally rotate more slowly, with periods of $P \sim 16$ ms to several s. 
In contrast, millisecond pulsars are significantly faster, with $P \sim 1.55$ ms to a few ms.
These rapid rotation rates result from the conservation of angular momentum during the collapse of their progenitor stars. 
As the stellar core contracts, angular momentum is redistributed, causing the newly formed neutron star to "spin up." 
Furthermore, it is believed that millisecond pulsars achieve these extreme rotation speeds through a spin-up process driven by the accretion of matter and angular momentum from a companion star in a binary system (see e.g. \cite{2001ASPC..248..469R} for a review).

The measurement of neutron star rotation is primarily based on electromagnetic observations. 
For instance, the precise timing of their pulsed emissions allows for an accurate determination of $P$. 
However, there are astrophysical scenarios in which no electromagnetic counterpart is observed, such as the binary neutron star merger event GW190425 \cite{2020ApJ...892L...3A}.
Additionally, this method of quantifying rotation by $P$ applies only to cases of rigid rotation. 
In astrophysical events like core-collapse supernovae and binary neutron star mergers, neutron stars are expected to exhibit differential rotation (see e.g. \cite{2007PhR...442...38J,2016nure.book.....S,2017RPPh...80i6901B} for reviews).

Neutron stars are also known to host the strongest magnetic fields in the Universe, with field strengths reaching up to $10^{14-15}$ G. 
Highly magnetized neutron stars, such as magnetars, are believed to explain several enigmatic astrophysical phenomena, including soft gamma-ray repeaters and anomalous X-ray pulsars \cite{1998Natur.393..235K,1999ApJ...510L.111H,1995ApJ...442L..17M,2000A&A...361..240M,1995A&A...299L..41V}. 
Moreover, magnetic fields can deform neutron stars, with the type of deformation depending on the field geometry. 
A purely toroidal magnetic field induces prolateness \cite{2008PhRvD..78d4045K,2009ApJ...698..541K,2012MNRAS.427.3406F}, whereas a purely poloidal field causes oblateness \cite{1995A&A...301..757B,2001A&A...372..594K,2012PhRvD..85d4030Y}. 
These deformations make rotating neutron stars potential sources of detectable continuous gravitational waves \cite{1996A&A...312..675B}. 
However, the true geometry of magnetic fields inside neutron stars remains uncertain. 
Stability analyses suggest that simple field configurations are prone to instabilities \cite{1957PPSB...70...31T,1973MNRAS.161..365T,1973MNRAS.163...77M,1974MNRAS.168..505M,1973MNRAS.162..339W}. Magnetohydrodynamic (MHD) simulations propose that a mixed toroidal-poloidal configuration, known as a "twisted torus," is the most favorable \cite{2006A&A...450.1077B,2006A&A...450.1097B,2009MNRAS.397..763B}. Nevertheless, recent studies (e.g. \cite{2023PhRvD.108h4006S}) suggest that instabilities may still persist in these mixed-field configurations.

Since the magnetic field governs neutron star emissions, understanding its configuration is critical. 
Surface magnetic field strengths are often estimated using the dipole spin-down model with radio observations \cite{2011MNRAS.411.2471C,2011MNRAS.415.1703C}. 
Recently, the Neutron Star Interior Composition Explorer (NICER) has enabled the deduction of the geometry and strength of surface magnetic fields from X-ray emitting hotspots in pulsars \cite{2019ApJ...887L..23B,2020ApJ...889..165D}.
However, these methods only provide information about surface fields, leaving the internal magnetic field largely unexplored. 
Determining the geometry and strength of magnetic fields inside neutron stars remains a significant challenge.

Gravitational wave detections offer a novel way to probe the internal properties of neutron stars. 
The first detected signal, GW170817, originated from a binary neutron star merger \cite{2017PhRvL.119p1101A}. 
Beyond mergers, violent astrophysical events like core-collapse supernovae and accretion-induced collapse could also excite oscillation modes in neutron stars, which are potential sources of detectable gravitational waves (see e.g. \cite{2004MNRAS.352.1089S,2006nwap.conf...25K}).
These oscillation modes are strongly tied to the composition, structure and dynamics of the stars, making their detection a valuable tool for studying neutron star interiors.
As mentioned, neutron stars are expected to possess both rapid rotation and strong magnetic fields, especially in these violent events that excite the oscillation modes.
For example, a magnetic field strength as high as $10^{17-18}~\mathrm{G}$ could even be reached in proto-neutron stars formed right after core-collapse supernovae \cite{2014MNRAS.439.3541P} and rotate with a period of $P \sim \mathcal{O}(1)~\mathrm{ms}$ \cite{2006ApJS..164..130O}.
Furthermore, binary neutron star merger simulations have demonstrated that the local maximum magnetic field can be amplified to $\sim 10^{17}~\mathrm{G}$ during the merger \cite{2006Sci...312..719P,2015PhRvD..92f4034K,2015PhRvD..92l4034K,2020PhRvD.102j3006A} and the hypermassive neutron stars formed from the mergers have a typical angular velocity of kHz (roughly corresponds to $P \sim \mathcal{O}(1)~\mathrm{ms}$) \cite{2017PhRvD..96d3004H}.
Thus, understanding the effects of rotation and magnetic fields on oscillation modes is essential for probing the internal properties of neutron stars in these events.

The oscillation modes of rotating neutron stars have been extensively studied using perturbative calculations and dynamical simulations \cite{1998MNRAS.299.1059A,1999ApJ...515..414Y,2001MNRAS.320..307K,2001MNRAS.325.1463F,2002ApJ...568L..41Y,2002PhRvD..65h4024F,2004PhRvD..70l4015B,2004MNRAS.352.1089S,2005MNRAS.356..217Y,2006MNRAS.368.1609D,2010PhRvD..81h4019K,2011PhRvD..83f4031G,2013PhRvD..88d4052D,2020PhRvL.125k1106K,2024arXiv240113993Y}. 
In particular, \cite{2004MNRAS.352.1089S} used dynamical simulations in the Cowling approximation (fixing spacetime while evolving matter equations) to study axisymmetric oscillation modes of uniformly and differentially rotating neutron stars and constructed linear relations for predicting the fundamental $l=0$ quasi-radial mode frequency $f_F$ and the fundamental $l=2$ quadrupolar mode frequency $f_{^2f}$ as functions of the ratio of kinetic energy to gravitational binding energy $T/|W|$.
Recently, our previous work \cite{2024arXiv240113993Y} updated these linear relations by including both various degrees of differential rotation and dynamical spacetime for the first time.
Hence, by measuring the fundamental mode frequency, we can infer the rotation of neutron stars, quantified by $T/|W|$, even in cases with differential rotation.

Similarly, oscillations in magnetized neutron stars have been studied using both Newtonian (e.g. \cite{10.1111/j.1365-2966.2010.17499.x}) and general relativistic approaches (e.g. \cite{2001MNRAS.328.1161M,2009MNRAS.395.1163S,2012MNRAS.421.2054G}).
Specifically, through dynamical simulations, our previous work \cite{2022CmPhy...5..334L} examined the axisymmetric oscillation modes of magnetized neutron stars with a purely toroidal field in dynamical spacetime and demonstrated the magnetic suppression effect on oscillation mode frequencies due to a strong toroidal field of $\sim 10^{17}$ G for the first time.
Moreover, by dynamically simulating the gravitational collapse of a magnetized neutron star due to a phase transition (typically known as phase-transition-induced collapse) for the first time \cite{2024MNRAS.534.3612Y}, we showed that the internal magnetic field strength of a neutron star can be directly constrained by measuring the frequency of excited fundamental modes in the corresponding gravitational wave signals generated during the collapse \cite{2025PhRvD.112d3035Y}.

Although the oscillations of rotating and magnetized neutron stars have been extensively studied individually, limited research has explored the combined effects of rapid rotation and strong magnetic fields. For instance, the oscillations of rotating magnetized neutron stars with purely toroidal magnetic fields \cite{2010MNRAS.405..318L} and purely poloidal magnetic fields \cite{2011MNRAS.412.1730L} have only been studied using a Newtonian framework.
However, this Newtonian approach is likely to introduce significant errors in determining the frequencies of oscillation modes. 
A study that incorporates both rotation and magnetic fields in dynamical spacetime is essential for improving our understanding of neutron star oscillations across different astrophysical scenarios.

To address this limitation, a promising approach would involve dynamical simulations capable of efficiently exploring a wide range of magnetized and rotating stellar models within a reasonable timeframe and with minimal computational resources.
\texttt{Gmunu} \citep{2020CQGra..37n5015C,2021MNRAS.508.2279C,2022ApJS..261...22C}, a new general-relativistic magnetohydrodynamics code, is well-suited for this purpose.
It supports simulations in multiple dimensions (1D, 2D, and 3D) and coordinate systems (cartesian, cylindrical, and spherical) through a block-based adaptive mesh refinement (AMR) module, enabling the imposition of symmetries to reduce dimensionality and significantly lower computational costs for various problems.
This flexibility enables users to choose the optimal dimensionality and coordinate system for their specific problem, and to impose symmetries that can further lower computational requirements when appropriate.
The 2D axisymmetric simulations conducted in this study are particularly well-suited for investigating axisymmetric ($m=0$) oscillation modes, such as the $F$-mode and $^2f$-mode studied by \cite{2004MNRAS.352.1089S,2006MNRAS.368.1609D,2024arXiv240113993Y}.
However, it should be emphasized that non-axisymmetric modes, including the $m=1$ and $m=2$ modes, may develop and play an important role during violent astrophysical events, such as binary neutron star mergers and core-collapse supernovae (see e.g. \cite{2015PhRvD..92l1502P}).
Therefore, by adopting 2D axisymmetry, this study is unable to capture such non-axisymmetric modes.
A comprehensive investigation of these modes would necessitate fully 3D simulations without imposing axisymmetry.
Additionally, \texttt{Gmunu} solves the elliptic metric equations under the conformally flat condition (CFC) approximation efficiently and robustly using the multigrid method.
These capabilities make \texttt{Gmunu} an excellent tool for studying neutron star problems involving rapid rotation and strong magnetic fields \cite{2021ApJ...915..108N,2022CmPhy...5..334L,2024MNRAS.534.3612Y,2025PhRvD.112d3035Y,Yip:2023qkh,2024arXiv240113993Y,2024ApJ...975..116C,2024PhRvD.110d3015C,2024PhRvD.110l4063M,2025PhRvD.111d3036C,2025PhRvD.111f3030C}.

In this work, for the first time, we account for both rotation and magnetic fields to investigate how the magnetic suppression effect on the fundamental mode frequency, identified in our previous work \cite{2022ApJ...934L..17R}, influences the linear relations quantifying $f_{F}$ and $f_{^2f}$ as we presented in \cite{2024arXiv240113993Y}.
Specifically, we begin by constructing initial rotating neutron star models with a purely toroidal magnetic field using the open-source code \texttt{XNS} \cite{2011A&A...528A.101B,2014MNRAS.439.3541P,2015MNRAS.447.2821P,2017MNRAS.470.2469P,2020A&A...640A..44S}.
These equilibrium models are then perturbed and evolved in dynamical spacetime using \texttt{Gmunu}.
The details of the initial neutron star models and their evolution are provided in Section \ref{sec:num_method}.
Next, in Sections \ref{sec:freq_compact} and \ref{sec:freq_ratio}, we investigate how the fundamental mode frequencies vary with the stellar compactness $M/R$ and the kinetic-to-binding energy ratio $T/|W|$ of neutron stars respectively.
Following this, we demonstrate that measuring the frequency ratio between the two fundamental modes $f_{^2f}/f_{F}$ can provide information about $\mathcal{B}_\mathrm{max}$ and the kinetic energy-to-binding ratio $T/|W|$ in Section \ref{sec:constrain}.
After that, we examine the deviations in the frequencies of the fundamental mode caused by differential rotation in Section \ref{sec_dr}, comparing them to the predictions derived from linear relations for uniformly rotating models.
Finally, the conclusions of this study are presented in Section \ref{sec:conclusions}.

Unless otherwise specified, we choose dimensionless units for the physical quantities by setting the speed of light, the gravitational constant, and the solar mass to one, $c=G=M_\odot=1$.

\section{Numerical methods}\label{sec:num_method}
\subsection{Initial neutron star models}
We construct equilibrium models of rotating neutron stars with a purely toroidal magnetic field in axisymmetry using the open-source code \texttt{XNS} \cite{2011A&A...528A.101B,2014MNRAS.439.3541P,2015MNRAS.447.2821P,2017MNRAS.470.2469P,2020A&A...640A..44S}. 
These models are used as initial data for our dynamical simulations.

A polytropic equation of state is adopted to compute initial neutron star models,
    \begin{equation}
    P=K \rho^\gamma,
    \end{equation}
where $P$ denotes the pressure, $\rho$ denotes the rest-mass density and we choose the polytropic constant $K =100$ and polytropic index $\gamma=2$. 

We assign the specific internal energy $\epsilon$ on the initial time-slice using,
    \begin{equation}
    \epsilon=\frac{K}{\gamma-1} \rho^{\gamma-1}.
    \end{equation}

A magnetic polytropic law is used to model the toroidal magnetic field \emph{enclosed in the star},
    \begin{equation}\label{eqn3}
        \mathcal{B}_{\phi}=\alpha^{-1}K_{\mathrm{m}}(\rho h\varpi^2)^m,
    \end{equation}
where $\alpha$ is the lapse function, $K_{\mathrm{m}}$ is the toroidal magnetization constant, $h$ is the specific enthalpy, $\varpi^2=\alpha^2\psi^4r^2\sin^2\theta$, $\psi$ is the conformal factor, $(r,\theta)$ is the radial and angular coordinates in 2D spherical coordinates, and $m\geq1$ is the toroidal magnetization index.

The toroidal magnetic field corresponds to a field aligned with the $\phi$-direction in spherical coordinates (see e.g. \cite{2017MNRAS.466.1330H} for a depiction of a toroidal magnetic field). 
As discussed in Section \ref{sec:intro}, purely toroidal field configurations are generally considered unstable. 
However, by employing 2D axisymmetry in our simulations, the magnetic instabilities associated with this configuration are suppressed. 
Nevertheless, many studies in the literature report instability even in mixed-field configurations (see e.g. \cite{2023PhRvD.108h4006S}).
As the stable magnetic field configuration remains uncertain, this work provides an initial investigation into the fundamental modes of rotating neutron stars with a purely toroidal magnetic field setup.

All models share the same toroidal magnetization index $m = 1$ but differ in the value of the toroidal magnetization constant, $K_{\mathrm{m}}$.
To investigate the effects of the toroidal magnetic field, we construct equilibrium models in 5 sequences with $K_{\mathrm{m}} \in \{0.5, 1.0, 1.5, 2.0, 2.5\}$, each corresponding to a maximum magnetic field strength of $\mathcal{B}_\mathrm{max} \sim \mathcal{O}(10^{17})$ G.
As mentioned in Section~\ref{sec:intro}, such a field strength has been shown to suppress the fundamental mode frequency of neutron stars in our previous work \cite{2022CmPhy...5..334L}.

Since this work does not aim to investigate neutron stars with varying masses, we adopt a fixed baryonic mass of $M_0 = 1.506$ for all models. Three high-mass neutron stars have been observed recently: J0348+0432 with $M = 2.01 \pm 0.04$ \cite{2013Sci...340..448A}, PSR J0740+6620 with $M = 2.08 \pm 0.07$ \cite{2021ApJ...915L..12F}, and PSR J0952-0607 with $M = 2.35 \pm 0.17$ \cite{2022ApJ...934L..17R}. 
These observations indicate that the maximum mass of a neutron star should be at least $M = 2.0$. 
Consequently, the masses of our models fall within the observational constraints on the maximum mass of a neutron star. 
Additionally, the adopted baryonic mass $M_0 = 1.506$ is identical to that used in previous studies \cite{2006MNRAS.368.1609D, 2024arXiv240113993Y}, allowing for direct comparison with these works.

By fixing the baryonic mass, we aim to clarify the qualitative trends in how mode frequencies vary with rotation and toroidal magnetic field strength. 
However, if the mass were not fixed but allowed to vary, both the mode frequencies and quantitative aspects, such as the slopes and intercepts of the linear fits discussed in Sections~\ref{sec:freq_compact} and \ref{sec:freq_ratio}, would be expected to change. 
Therefore, a more comprehensive study that explores a range of baryonic masses, including values close to $2.0$, which are especially relevant for binary neutron star merger remnants, will be necessary for a more general determination of mode frequencies. 
Such an analysis is beyond the scope of this study and will be addressed in future work.

The detailed properties of the equilibrium models used in this study are summarized in Appendix~\ref{sec_equil_models}.

\subsection{Evolutions}
We use the new general relativistic magnetohydrodynamics code \texttt{Gmunu} \cite{2020CQGra..37n5015C,2021MNRAS.508.2279C,2022ApJS..261...22C} to evolve the stellar models in dynamical spacetime.
The models are evolved over a time of 20 ms using a polytropic equation of state, $P = K \rho^\gamma$, with the same parameters as the equilibrium models (i.e. $K = 100$ and $\gamma = 2$).

We perform 2D ideal general-relativistic magnetohydrodynamics (GRMHD) simulations in axisymmetry about the $z$-axis, assuming equatorial symmetry and employing spherical coordinates $(r, \theta)$.
The computational domain spans $0 \leq r \leq 60$ and $0 \leq \theta \leq \pi/2$, with a base grid resolution of $N_r \times N_\theta = 64 \times 16$. 
AMR with 4 levels is used, achieving an effective resolution of $512 \times 128$. 
The AMR refinement criteria follow those described in \cite{2021MNRAS.508.2279C,2022CmPhy...5..334L}.
Our simulations employ a total variation diminishing Lax-Friedrichs (TVDLF) approximate Riemann solver \cite{1996JCoPh.128...82T}, a third-order reconstruction method using the piecewise parabolic method (PPM) \cite{1984JCoPh..54..174C} and a third-order accurate strong stability-preserving Runge-Kutta (SSPRK3) time integrator \cite{1988JCoPh..77..439S}. 
An artificial atmosphere with a rest-mass density $\rho_\mathrm{atm} \sim 10^{-10} \rho_\mathrm{c}$ is imposed outside the star.
Furthermore, no divergence cleaning method is applied in these simulations since they are restricted to magnetized stars with a purely toroidal magnetic field in axisymmetry.

\subsection{Initial perturbations}\label{sec:perturbations}
By applying the following fluid perturbations in the initial time-slice,
we excite the 2 fundamental modes of neutron stars \cite{2006MNRAS.368.1609D}.

First, the fundamental $l=0$ quasi-radial mode (i.e. ${F}$-mode) is excited by the $l=0$ perturbation on the $r$-component of the three-velocity field $v^r$,
    \begin{equation}
        \delta v^r=a \sin\left[\pi\frac{r}{r_{\mathrm{s}}(\theta)}\right],
    \end{equation}
where $r_{\mathrm{s}}(\theta)$ denotes the radial position of the stellar surface, and the perturbation amplitude $a$ (in the unit of $c$) is chosen to be -0.005.

Second, the fundamental $l=2$ quadrupolar mode (i.e. ${^2f}$-mode) is excited by the $l=2$ perturbation on the $\theta$-component of the three-velocity field $v^{\theta}$,
    \begin{equation}
        \delta v^{\theta}=a \sin\left[\pi\frac{r}{r_{\mathrm{s}}(\theta)}\right] \sin\theta \cos\theta,
    \end{equation}
where $a$ is chosen to be 0.01.

We extract the fundamental modes by performing Fourier transforms of $v^r$ and $v^\theta$ at $r = 3$ and $\theta = \pi/4$, ensuring that the extraction position lies within the star (see \cite{2006MNRAS.368.1609D} for further details).

\section{Fundamental mode frequencies against stellar compactness}\label{sec:freq_compact}
\begin{figure*}[htbp]
    \centering
    \includegraphics[width=\textwidth, angle=0]{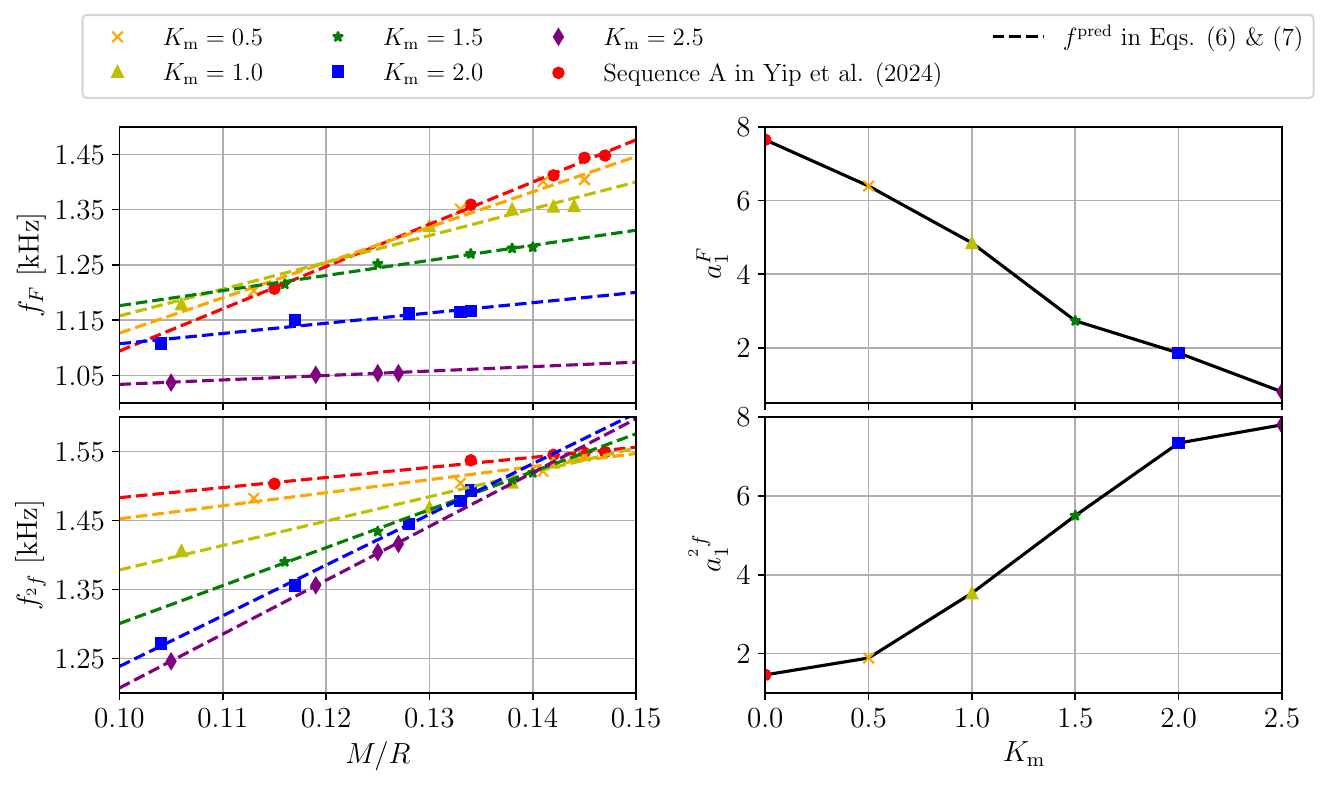}
    \caption{
             Plots of fundamental mode frequencies $f$ against the stellar compactness $M/R$ (left panels), where $M$ is the gravitational mass and $R$ is the circumferential radius, and the corresponding slopes $a_1$ of the linear fits $f^\mathrm{pred}(M/R)$ against the toroidal magnetization constant $K_{\mathrm{m}}$ (right panels).
             Specifically, we plot the fundamental $l=0$ quasi-radial mode frequency $f_{F}$ (top left panel) and the fundamental $l=2$ quadrupolar mode frequency $f_{^2f}$ (bottom left panel) against $M/R$.
             The corresponding slopes of the linear fits $a^F_1$ (top right panel) and $a^{^2f}1$ (bottom right panel) are plotted against $K_{\mathrm{m}}$.
             Data points represent five sequences with $K_{\mathrm{m}} \in \{0.5, 1.0, 1.5, 2.0, 2.5\}$, where $K_{\mathrm{m}}$ quantifies the strength of the toroidal magnetic field.
             For comparison, we also include data from sequence A of Yip et al. \cite{2024arXiv240113993Y}, which investigates the fundamental modes of unmagnetized neutron stars undergoing uniform rotation.
             As in the unmagnetized case (sequence A), both $f_{F}$ and $f_{^2f}$ increase approximately linearly with $M/R$ for all values of $K_{\mathrm{m}}$.
             Linear regressions are performed to obtain the predicted linear fit $f^\mathrm{pred}(M/R)$ in the form of Eqs.~(\ref{eqn6}) and (\ref{eqn7}) for each sequence (dashed lines).
             We observe that the slopes $a^F_1$ and $a^{^2f}_1$ deviate from the unmagnetized case: $a^F_1$ decreases with $K_{\mathrm{m}}$ while $a^{^2f}_1$ increases with $K_{\mathrm{m}}$.
             Therefore, this demonstrates that the quasi-linear relation between the fundamental mode frequency and stellar compactness remains valid for rotating neutron stars in the presence of a toroidal magnetic field, although the field strength, quantified by $K_{\mathrm{m}}$, modifies the slope of the relation.
            }
    \label{fig1}	
\end{figure*}

Our previous work have shown the quasi-linearity between the fundamental mode frequency and the stellar compactness for rotating neutron stars without a magnetic field \cite{2024arXiv240113993Y}.
To determine whether the presence of a toroidal magnetic field affects this relation, we plot the fundamental $l=0$ quasi-radial mode frequency $f_{F}$ (top left panel) and fundamental $l=2$ quadrupolar mode frequency $f_{^2f}$ (bottom left panel) against the stellar compactness $M/R$ in Fig.~\ref{fig1}, where $M$ is the gravitational mass and $R$ is the circumferential radius.
The data points are arranged into 5 sequences with $K_{\mathrm{m}} \in \{0.5, 1.0, 1.5, 2.0, 2.5\}$, where $K_{\mathrm{m}}$ is the toroidal magnetization constant quantifying the toroidal magnetic field strength.
As a comparison, we also include data from sequence A of Yip et al. \cite{2024arXiv240113993Y}, which investigates the fundamental modes of unmagnetized neutron stars undergoing uniform rotation.
As in the unmagnetized case (sequence A), both $f_{F}$ and $f_{^2f}$ increase approximately linearly with $M/R$ for all values of $K_{\mathrm{m}}$.

Hence, we perform linear regressions using the data from our simulations and sequence A to quantify $f_{F}$ and $f_{^2f}$ as functions of the stellar compactness $M/R$ for rotating neutron stars with a purely toroidal magnetic field.
For the quasi-radial $l=0$ fundamental mode frequency $f^\mathrm{pred}_{F}$,
\begin{equation}\label{eqn6}
f^\mathrm{pred}_{F}(\mathrm{kHz}) \approx a^{F}_0 - a^{F}_1\frac{M}{R},
\end{equation}
where the residues of $f^\mathrm{pred}_{F}(M/R)$ are $\lesssim 1\%$.
For the quadrupolar $l=2$ fundamental mode frequency $f^\mathrm{pred}_{^2f}$,
\begin{equation}\label{eqn7}
f^\mathrm{pred}_{^2f}(\mathrm{kHz}) \approx a^{^2f}_0 - a^{^2f}_1\frac{M}{R},
\end{equation}
where the residues of $f^\mathrm{pred}_{^2f}(M/R)$ are $\lesssim 1\%$.

To better illustrate how the slopes of the linear fits $f^\mathrm{pred}(M/R)$ change with $K_{\mathrm{m}}$, we plot $a^F_1$ (top right panel) and $a^{2f}_1$ (bottom right panel) against the toroidal magnetization constant $K_{\mathrm{m}}$ in Fig.~\ref{fig1}.
We observe that the slopes $a^F_1$ and $a^{2f}_1$ deviate from the unmagnetized case. 
Specifically, $a^F_1$ decreases with increasing $K_{\mathrm{m}}$, whereas $a^{2f}_1$ increases with $K_{\mathrm{m}}$.
Hence, this demonstrates that the quasi-linear relation between the fundamental mode frequency and stellar compactness remains valid for rotating neutron stars in the presence of a toroidal magnetic field, although the field strength, quantified by $K_{\mathrm{m}}$, modifies the slope of the relation.

%

%

\section{Fundamental mode frequencies against kinetic-to-binding energy ratio}\label{sec:freq_ratio}
\begin{figure*}[htbp]
    \centering
    \includegraphics[width=\textwidth, angle=0]{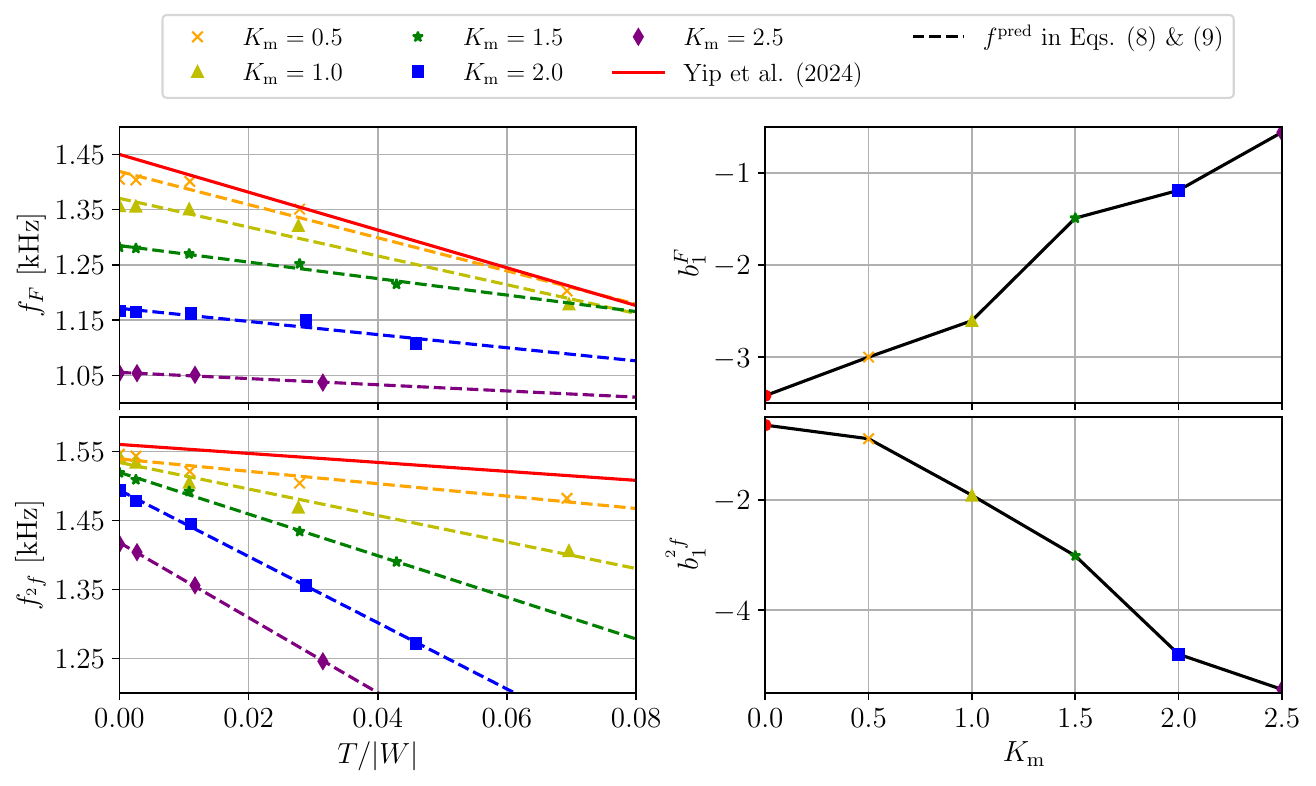}
    \caption{Plots of fundamental mode frequencies $f$ against the kinetic-to-binding energy ratio $T/|W|$ (left panels) and the corresponding slopes $b_1$ of the linear fits $f^\mathrm{pred}(T/|W|)$ against the toroidal magnetization constant $K_{\mathrm{m}}$ (right panels).
    Specifically, we plot the fundamental $l=0$ quasi-radial mode frequency $f_{F}$ (top left panel) and the fundamental $l=2$ quadrupolar mode frequency $f_{^2f}$ (bottom left panel) against $T/|W|$.
    The corresponding slopes of the linear fits $b^F_1$ (top right panel) and $b^{^2f}_1$ (bottom right panel) are plotted against $K_{\mathrm{m}}$.
    The data points are grouped into five sequences with $K_{\mathrm{m}} \in \{0.5, 1.0, 1.5, 2.0, 2.5\}$, where $K_{\mathrm{m}}$ quantifies the toroidal field strength.
    Both $f_{F}$ and $f_{^2f}$ increase approximately linearly with $T/|W|$ for all $K_{\mathrm{m}}$. 
    Accordingly, linear regressions are performed to obtain the predicted linear fit $f^\mathrm{pred}\left(T/|W|\right)$ in the form of Eqs. (\ref{eqn8}) and (\ref{eqn9}) for each sequence (dashed lines).
    We also compare with the predictions $f^\mathrm{pred}_{F}$ and $f^\mathrm{pred}_{^2f}$ in Eqs. (4) and (5) of our previous work of Yip et al. \cite{2024arXiv240113993Y} (red solid lines). 
    This prediction was based on the fundamental modes of rotating neutron stars with differential rotation but without magnetic fields.
    We observe that the slopes $b^F_1$ and $b^{^2f}_1$ deviate from the unmagnetized case (Yip et al.): $b^F_1$ increases with $K_{\mathrm{m}}$ while $b^{^2f}_1$ decreases with $K_{\mathrm{m}}$.
    Therefore, the quasi-linear relation between the fundamental mode frequency and the kinetic-to-binding energy ratio still holds for rotating neutron stars when a toroidal magnetic field is present, though the field strength, quantified by $K_{\mathrm{m}}$, modifies the slope of the relation.}
    \label{fig2}	
\end{figure*}

As mentioned in Section \ref{sec:intro}, our previous work has updated the linear relations for quantifying fundamental $l=0$ quasi-radial mode frequency $f_{F}$ and fundamental $l=2$ quadrupolar mode frequency $f_{^2f}$ of rotating neutron stars based on the kinetic-to-binding energy ratio $T/|W|$ by considering both the effect of dynamical spacetime and various degrees of differential rotation for the first time \cite{2024arXiv240113993Y}.
Nevertheless, a magnetic field is not taken into account in constructing the relations.
Hence, we revisit the relation between the fundamental mode frequency $f$ and kinetic-to-binding energy ratio $T/|W|$ by considering the presence of a toroidal magnetic field.
In Fig~\ref{fig2}, we plot the frequency of the fundamental $l=0$ quasi-radial mode $f_{F}$ (top left panel) and the fundamental $l=2$ quadrupolar mode $f_{^2f}$ (bottom left panel) against the kinetic-to-binding energy ratio $T/|W|$.
The data points are grouped into five sequences with $K_{\mathrm{m}} \in \{0.5, 1.0, 1.5, 2.0, 2.5\}$.
For all $K_{\mathrm{m}}$ values, $f_{F}$ increases approximately linearly with $T/|W|$. 

Thus, we perform linear regressions using our simulation data to obtain a linear fit for each sequence to quantify the fundamental $l=0$ quasi-radial mode frequency $f_{F}$ and fundamental $l=2$ quadrupolar mode frequency $f_{^2f}$ as functions of the kinetic-to-binding energy ratio $T/|W|$ for rotating neutron stars with a purely toroidal magnetic field.
For the quasi-radial $l=0$ fundamental mode frequency $f^\mathrm{pred}_{F}$,

    \begin{equation}\label{eqn8}
        f^\mathrm{pred}_{F}(\mathrm{kHz}) \approx b^{F}_0 - b^{F}_1\frac{T}{|W|},
    \end{equation} 
where the residues of $f^\mathrm{pred}_{F}(T/|W|)$ are $\lesssim 2\%$.
For the quadrupolar $l=2$ fundamental mode frequency $f^\mathrm{pred}_{^2f}$,
    \begin{equation}\label{eqn9}
        f^\mathrm{pred}_{^2f}(\mathrm{kHz}) \approx b^{^2f}_0 - b^{^2f}_1\frac{T}{|W|},
    \end{equation}
where the residues of $f^\mathrm{pred}_{^2f}(T/|W|)$ are $\lesssim 1\%$.
We also include the relations proposed in Eqs. (4) and (5) of our previous work \cite{2024arXiv240113993Y} (red solid line).
For the quasi-radial $l=0$ fundamental mode frequency $f^\mathrm{pred}_{F}$,
        \begin{equation}\label{eqn10}
            f^\mathrm{pred}_{F}(\mathrm{kHz}) \approx 1.45-3.42 \frac{T}{|W|}.
        \end{equation}        
For the quadrupolar $l=2$ fundamental mode frequency $f^\mathrm{pred}_{^2f}$,
        \begin{equation}\label{eqn11}
            f^\mathrm{pred}_{^2f}(\mathrm{kHz}) \approx 1.56-0.65 \frac{T}{|W|}.
        \end{equation}       
        
To better illustrate how the slopes of the linear fits $f^\mathrm{pred}(T/|W|)$ change with $K_{\mathrm{m}}$, we plot $b^F_1$ (top right panel) and $b^{2f}_1$ (bottom right panel) against the toroidal magnetization constant $K_{\mathrm{m}}$ in Fig.~\ref{fig2}.
We observe that the slopes $b^F_1$ and $b^{^2f}_1$ deviate from the unmagnetized case (Yip et al.).
In particular, $b^F_1$ increases with $K_{\mathrm{m}}$ while $b^{^2f}_1$ decreases with $K_{\mathrm{m}}$.
Accordingly, the quasi-linear relation between the fundamental mode frequency and the kinetic-to-binding energy ratio still holds for rotating neutron stars when a toroidal magnetic field is present, though the field strength, quantified by $K_{\mathrm{m}}$, modifies the slope of the relation.

%
%
%

\section{\label{sec:constrain}Constraining the magnetic field and the rotation}
\begin{figure}[htbp]
	\centering
	\includegraphics[width=\columnwidth, angle=0]{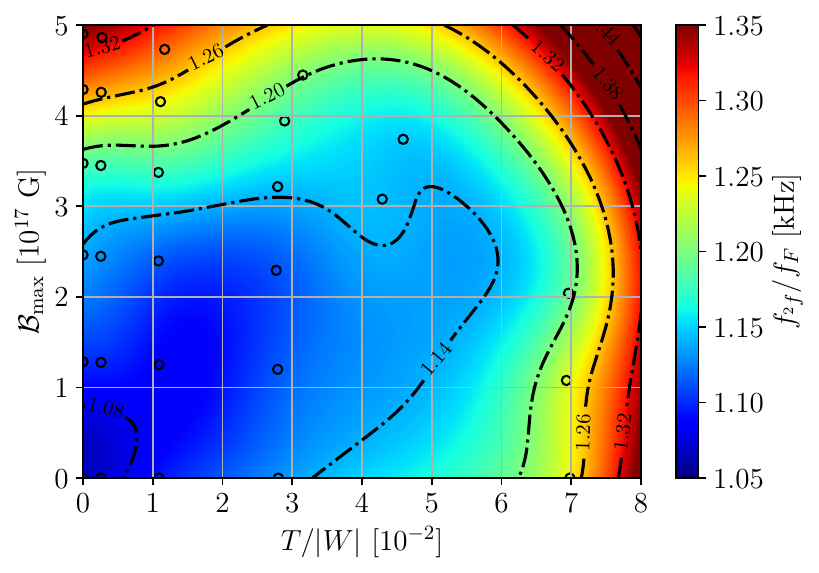}
	\caption{\label{fig3} 
             Contour plot of the frequency ratio between the fundamental $l=0$ quasi-radial mode and the fundamental $l=2$ quadrupolar mode $f_{^2f}/f_{F}$ against the kinetic-to-binding energy ratio $T/|W|$ (horizontal axis) and the maximum magnetic field strength $\mathcal{B}_\mathrm{max}$ (vertical axis).
             This plot is constructed by the multiquadric radial basis function interpolation of the data points of our models and the models in Sequence A of Yip et al. \cite{2024arXiv240113993Y}.
             Each data point is labeled by a colored dot with a black edge.
             The dash-dotted lines denote the contour lines for particular values of $f_{^2f}/f_{F}$.
             Values of $f_{^2f}/f_{F}$ are located in specific regions of the $\mathcal{B}_\mathrm{max}$ - $T/|W|$ plane.
             In consequence, measuring $f_{^2f}/f_{F}$ allows one to infer the value of $\mathcal{B}_\mathrm{max}$ and $T/|W|$.
	}
\end{figure}

To better illustrate the correlation between the fundamental mode frequencies and the properties of the rotating magnetized neutron star, we plot a contour plot of the frequency ratio between the fundamental $l=0$ quasi-radial mode and the fundamental $l=2$ quadrupolar mode $f_{^2f}/f_{F}$ against the kinetic-to-binding energy ratio $T/|W|$ (horizontal axis) and the maximum magnetic field strength $\mathcal{B}_\mathrm{max}$ (vertical axis) in Fig.~\ref{fig3}.
We constructed this plot is constructed by the multiquadric radial basis function interpolation of the data points of our models and the models in Sequence A of Yip et al. \cite{2024arXiv240113993Y}.
A colored dot with a black edge labels each data point.
The dash-dotted lines denote the contour lines for particular values of $f_{^2f}/f_{F}$.
We use $T/|W|$ to quantify rotation instead of the rotational angular velocity $\Omega$ because $T/|W|$ can be generalized to cases with differential rotation, where no unique value of $\Omega$ is sufficient to describe the rotation. 
The effects of differential rotation are further discussed in Section \ref{sec_dr}. 
For a fixed value of $f_{^2f}/f_{F}$, there are localized regions in the $T/|W|$ - $\mathcal{B}_\mathrm{max}$ plane.
Therefore, the measurement of $f_{^2f}/f_{F}$ constrains the values of $T/|W|$ and $\mathcal{B}_\mathrm{max}$.

\section{Impact of differential rotation}\label{sec_dr}
As discussed in Section \ref{sec:intro}, neutron stars could exhibit differential rotation in some astrophysical scenarios, such as core-collapse supernovae and binary neutron star mergers (see e.g. \cite{2007PhR...442...38J,2016nure.book.....S,2017RPPh...80i6901B} for reviews).
In this section, we investigate the deviations caused by differential rotation in comparison to the fundamental mode frequency predicted by the linear relations $f^\mathrm{pred}\left(T/|W|\right)$ obtained for uniformly rotating cases in Section \ref{sec:freq_ratio}.

A variety of differential rotation laws have been proposed in the literature to describe the differential rotation of protoneutron stars and binary neutron star merger remnants (see e.g \cite{2017PhRvD..96d3004H,2017PhRvD..96j3011U,2022MNRAS.510.2948I,2024MNRAS.532..945C}).
However, since the primary focus of this section is on the effects of the degree of differential rotation rather than the detailed differences among rotation laws, we adopt a simplified case of the $j$-constant differential rotation law \cite{1989MNRAS.237..355K,1989MNRAS.239..153K}, which models differential rotation in neutron stars as,

\begin{equation}
j(\Omega) = A^2\left(\Omega_{\mathrm{c}} - \Omega\right),
\end{equation}
where $j$ is the relativistic specific angular momentum, $A$ is a parameter controlling the degree of differential rotation, $\Omega$ is the angular velocity, and $\Omega_\mathrm{c}$ is the central angular velocity.

The $j$-constant law is widely used to model differentially rotating neutron stars (e.g.\cite{2000ApJ...528L..29B,2004ApJ...610..941M,2014ApJ...790...19K}). Following previous studies (e.g. \cite{2010PhRvD..81h4019K,2024arXiv240113993Y}), we adopt the parameter $\Tilde{A} = (A/r_\mathrm{e})^{-1}$ to quantify the degree of differential rotation, where $r_\mathrm{e}$ is the equatorial radius of the equilibrium model.
As we do not intend to conduct a comprehensive survey of the effects of $\Tilde{A}$ and $K_\mathrm{m}$, we fix $\Tilde{A} = 1$ and select two representative, extreme values of $K_{\mathrm{m}} \in {0.5, 2.5}$ for this study. 
The adopted values of $K_{\mathrm{m}}$ correspond to the lowest and highest toroidal magnetization constants considered, thereby allowing us to estimate the impact of differential rotation across the range of magnetic field strengths relevant to our models. 
The choice of $\Tilde{A} = 1$ is motivated by its widespread use in studies of oscillations in differentially rotating neutron stars (see e.g. \cite{2004MNRAS.352.1089S, 2006MNRAS.368.1609D}) and it is physically representative because it yields a degree of differential rotation similar to that typically observed in core-collapse supernova simulations (see e.g. \cite{2004A&A...418..283V}).
Therefore, our parameter choices are representative of physically relevant scenarios and enable us to capture the essential effects of differential rotation and magnetic fields within a manageable computational framework.

\begin{figure*}[htbp]
    \centering
    \includegraphics[width=\textwidth, angle=0]{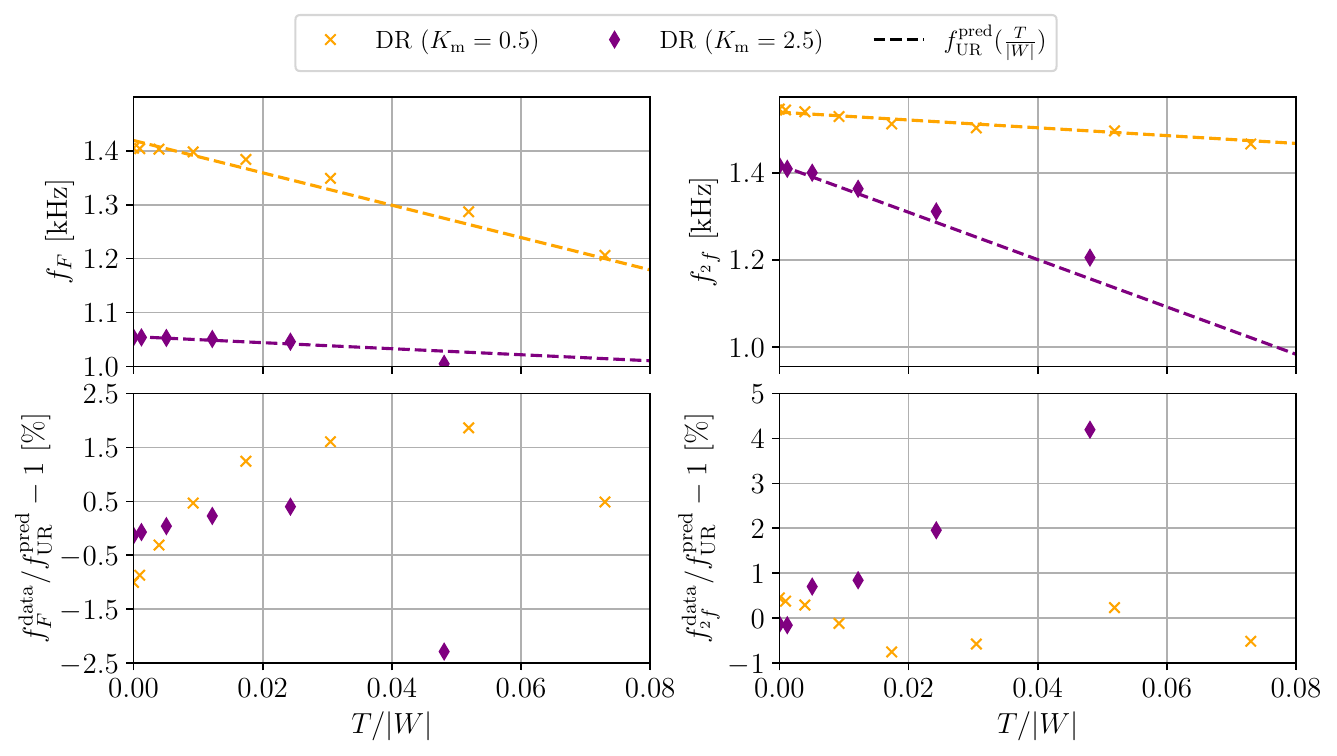}
    \caption{\label{fig4}
    Plots of fundamental mode frequencies $f$ (top panels) and the frequency deviations between the simulation data $f^\mathrm{data}$ of differentially rotating (DR) neutron star and the predictions by linear fits constructed with the data of uniformly rotating (UR) neutron star $f^\mathrm{pred}_\mathrm{UR}$ (bottom panels) against kinetic-to-binding energy ratio $T/|W|$.
    In particular, we plot fundamental $l=0$ quasi-radial mode frequency $f_{F}$ (top left panel), deviation of $f_{F}$ (bottom left panel), fundamental $l=2$ quadrupolar mode frequency $f_{^2f}$ (top right panel), and deviation of $f_{^2f}$ (bottom right panel) against $T/|W|$.
    The data points are grouped into 2 sequences of DR neutron star with $K_{\mathrm{m}} \in \{0.5, 2.5\}$.
    We find that only a slight deviation between $f^\mathrm{data}$ and $f^\mathrm{pred}_\mathrm{UR}$ with $f^\mathrm{data}_{F}/f^\mathrm{pred}_{F} - 1 \lesssim 2.5 \%$ and $f^\mathrm{data}_{^2f}/f^\mathrm{pred}_{^2f} - 1 \lesssim 5 \%$.
    Thus, we found that differential rotation in the examined models introduces only minor deviations in the fundamental mode frequencies compared to the predictions from linear relations derived for uniformly rotating cases.}
\end{figure*}

In Fig.~\ref{fig4}, we plot fundamental mode frequencies $f$ (top panels) and the frequency deviations between the simulation data $f^\mathrm{data}$ of differentially rotating (DR) neutron star and the predictions by linear fits constructed with the data of uniformly rotating (UR) neutron star $f^\mathrm{pred}_\mathrm{UR}$ (bottom panels) against kinetic-to-binding energy ratio $T/|W|$.
In particular, we plot fundamental $l=0$ quasi-radial mode frequency $f_{F}$ (top left panel), deviation of $f_{F}$ (bottom left panel), fundamental $l=2$ quadrupolar mode frequency $f_{^2f}$ (top right panel), and deviation of $f_{^2f}$ (bottom right panel) against $T/|W|$.
The data points are grouped into 2 sequences of DR neutron star with $K_{\mathrm{m}} \in \{0.5, 2.5\}$.
We find that only a slight deviation between $f^\mathrm{data}$ and $f^\mathrm{pred}_\mathrm{UR}$ with $f^\mathrm{data}_{F}/f^\mathrm{pred}_{F} - 1 \lesssim 2.5 \%$ and $f^\mathrm{data}_{^2f}/f^\mathrm{pred}_{^2f} - 1 \lesssim 5 \%$.
Consequently, differential rotation in the examined models introduces only minor deviations in the fundamental mode frequencies compared to the predictions from linear relations derived for uniformly rotating cases. 

Adopting a simplified $j$-constant differential rotation law with a fixed value of $\Tilde{A} = 1$ and representative values of $K_{\mathrm{m}}$ is expected to capture the qualitative trend of the quasi-linear behavior observed in the mode frequencies of magnetized neutron stars with differential rotation. 
However, this assumption may result in quantitative differences in the computed mode frequencies and, consequently, in deviations from the predicted linear fits, particularly for configurations outside the explored parameter space or those with more realistic rotation laws.
Therefore, a more comprehensive exploration of the parameter space of $\Tilde{A}$ and $K_{\mathrm{m}}$, together with the adoption of more realistic rotation laws, is necessary to determine the precise range of validity of these findings.

\section{Conclusions}\label{sec:conclusions}
In this work, for the first time, we considered a magnetic field to construct linear relations quantifying the frequencies of fundamental $l=0$ quasi-radial mode $f_{F}$ and fundamental $l=2$ quadrupolar mode $f_{^2f}$.
In particular, through linear regression of our simulated data, we found that the quasi-linearity between fundamental mode frequency $f$, stellar compactness $M/R$, the ratio of kinetic energy to binding energy $T/|W|$ remains valid but the slope is controlled by the toroidal magnetization constant $K_{\mathrm{m}}$ quantifying the toroidal magnetic field strength.
Next, we demonstrated that measuring the frequency ratio between the two fundamental modes $f_{^2f}/f_{F}$ can infer the kinetic-to-binding energy ratio $T/|W|$ and the maximum magnetic field strength $\mathcal{B}_\mathrm{max}$.
Furthermore, we examined 2 sequences of differentially rotating neutron star models with $K_{\mathrm{m}} \in \{0.5, 2.5\}$ and show that differntial rotations only introduces minor deviations in the fundamental mode frequencies compared to the predictions from linear relations derived for uniformly rotating cases in the examined models.

The fundamental modes discussed in this study lie in the frequency range of $f \sim 1200 - 1600$ Hz. 
Current gravitational wave detectors, such as Advanced LIGO \cite{2015CQGra..32g4001L}, Advanced Virgo \cite{2015CQGra..32b4001A}, and KAGRA \cite{2012CQGra..29l4007S,2013PhRvD..88d3007A}, are sensitive to gravitational wave signals within a frequency range of $f \sim 20 - 2000$ Hz. 
Consequently, these detectors can only barely detect the modes in this range.
On the other hand, third-generation gravitational wave detectors, including the Einstein Telescope (ET) \cite{2010CQGra..27s4002P} and Cosmic Explorer (CE) \cite{2017CQGra..34d4001A,2019BAAS...51c.141R,2019BAAS...51g..35R}, are designed for a significantly broader sensitivity band, covering frequencies from $f \sim 1 - 10000$ Hz. 
Moreover, it has been shown that the dominant mode frequency of the post-merger signal from binary neutron star coalescences at a distance of 68 Mpc can be measured with an accuracy of $\sim \mathcal{O}(10)$ Hz for post-merger signals with a signal-to-noise ratio of $\sim 10$ using third-generation detectors \cite{2023PhRvD.107l4009P}.
Assuming that the fundamental modes studied here can be measured with similar accuracy, and considering that the mode frequencies lie within the range $f \sim 1200 - 1600$ Hz while the frequency ratio spans $f_{^2f}/f_{F} \sim 1.05$–$1.35$ (as shown in Fig. \ref{fig3}), we expect that the frequency ratio $f_{^2f}/f_{F}$ can be determined with an error of $\lesssim 1\%$.
This measurement uncertainty would translate to an error bar of comparable width ($\lesssim 1\%$) when inferring the kinetic-to-binding energy ratio $T/|W|$ and the maximum magnetic field strength $\mathcal{B}_\mathrm{max}$ with the contour plot in Fig.~\ref{fig3}.
However, a thorough analysis focused on gravitational waves emitted by rotating neutron stars is required to confirm this claim. 
Such an investigation falls outside the scope of the present study and is left for future work.

Several extensions to the current work could be made. 
Firstly, a broader investigation covering a wide range of baryonic masses, including those approaching $2.0$, which are particularly relevant for merger remnants of binary neutron stars, would be beneficial to more generally characterize mode frequencies.
Additionally, for a more precise determination of oscillation modes in neutron stars, especially in contexts such as core-collapse supernovae and neutron star mergers, it is crucial to implement more realistic equations of state that account for thermal effects (see e.g. \cite{2017RPPh...80i6901B,2021PhRvD.103f3014C}) and to consider empirically motivated rotation profiles (see e.g. \cite{2017PhRvD..96d3004H,2017PhRvD..96j3011U,2022MNRAS.510.2948I,2024MNRAS.532..945C,2024PhRvD.110d3015C,2024PhRvD.110l4063M}), as have been proposed for modeling such scenarios.
Moreover, it would be worthwhile to explore alternative magnetic field configurations, including purely poloidal fields and twisted torus structures.
Finally, since the present work restricts itself to axisymmetry, thereby suppressing the instability associated with purely toroidal fields, future studies should include fully 3D simulations without this symmetry constraint. 
This would allow investigation of the stability and the emergence of non-axisymmetric oscillation modes, which could play a crucial role in the dynamics of binary neutron star remnants and proto-neutron stars formed in core-collapse supernovae (see e.g. \cite{2015PhRvD..92l1502P}).

\begin{acknowledgments}
We acknowledge the support of the CUHK Central High-Performance Computing Cluster, on which the simulations in this work have been performed. 
This work was partially supported by grants from the Research Grants Council of Hong Kong (Project No. CUHK 14306419), the Croucher Innovation Award from the Croucher Foundation Hong Kong, and the Direct Grant for Research from the Research Committee of The Chinese University of Hong Kong. 
\end{acknowledgments}

\appendix
\section{Equilibrium models of neutron stars}\label{sec_equil_models}
This appendix summarizes the detailed properties of the equilibrium models used in this study. 
Tables~\ref{tablea1} – \ref{tablea5} present the stellar properties of the equilibrium models for the sequences TK1U, TK2U, TK3U, TK4U, and TK5U.
These sequences represent uniformly rotating neutron stars with a fixed rest mass of $M_{0} = 1.506$ and toroidal magnetization constants $K_{\mathrm{m}} \in \{0.5, 1.0, 1.5, 2.0, 2.5\}$, respectively.
The data in these tables were discussed and used to construct the linear fits described in Sections \ref{sec:freq_compact} and \ref{sec:freq_ratio} of the main text.
Tables~\ref{tablea6}–~\ref{tablea7} summarize the properties of the equilibrium models for the sequences TK1D and TK5D. 
These sequences represent differentially rotating neutron stars modeled by the $j$-constant law with $\Tilde{A}=1.0$, a fixed rest mass of $M_{0} = 1.506$, and toroidal magnetization constants $K_{\mathrm{m}} \in \{0.5, 2.5\}$, respectively.
The data are provided in the form of one table for each sequence. Each table lists the model, the central rest mass density $\rho_\mathrm{c}$, the gravitational mass $M$, the circumferential radius $R$, the maximum magnetic field strength $\mathcal{B}_\mathrm{max}$, the central angular velocity $\Omega_\mathrm{c}$, and the ratio of rotational kinetic energy $T$ to the absolute value of gravitational binding energy $T/|W|$.

\setcounter{table}{0}
\renewcommand{\thetable}{A\arabic{table}}

\begin{table}[H]
	\centering
	\begin{tabular}{ccccccc}
		 Model & $\rho_\mathrm{c}$ & $M$ & $R$ & $\mathcal{B}_\mathrm{max}$ & $\Omega_\mathrm{c}$ & $T/|W|$\\
        & ($10^{-3}$) &  &   & ($10^{17}$ G)  & ($10^{-2}$)  &    \\
		\hline
  	 TK1U0 & 1.290 & 1.401 & 9.638  & 1.285 & 0.000 & 0.000 \\
		 TK1U1 & 1.275 & 1.402 & 9.687  & 1.277 & 0.500 & 0.003 \\
		 TK1U2 & 1.225 & 1.403 & 9.972  & 1.252 & 1.000 & 0.011 \\
		 TK1U3 & 1.130 & 1.407 & 10.544 & 1.200 & 1.500 & 0.028 \\
		 TK1U4 & 0.924 & 1.415 & 12.484 & 1.078 & 2.000 & 0.069 \\
	\end{tabular}
	\caption{\label{tablea1} Stellar properties of the equilibrium models of sequence TK1U constructed by the \texttt{XNS} code.
	TK1U is a sequence of uniformly rotating neutron star with a fixed rest mass $M_{0} = 1.506$ and toroidal magnetization constant $K_{\mathrm{m}}=0.5$.
	$\rho_\mathrm{c}$ is the central density, $M$ is the gravitational mass, $R$ is the circumferential radius, $\mathcal{B}_\mathrm{max}$ is the maximum magnetic field strength, $\Omega_\mathrm{c}$ is the central angular velocity, $T/|W|$ is the ratio of rotational kinetic energy $T$ to the absolute value of gravitational binding energy $|W|$.
    Numerical values of these stellar properties are rounded off to three decimal places.}
\end{table}

\begin{table}[H]
	\centering
	\begin{tabular}{ccccccc}
		 Model & $\rho_\mathrm{c}$ & $M$ & $R$ & $\mathcal{B}_\mathrm{max}$ & $\Omega_\mathrm{c}$ & $T/|W|$\\
        & ($10^{-3}$) &  &   & ($10^{17}$ G)  & ($10^{-2}$)  &    \\
		\hline
  	 TK2U0 & 1.316 & 1.404 & 9.773  & 2.464 & 0.000 & 0.000 \\
		 TK2U1 & 1.301 & 1.404 & 9.868  & 2.448 & 0.500 & 0.003 \\
		 TK2U2 & 1.253 & 1.406 & 10.154 & 2.397 & 1.000 & 0.011 \\
		 TK2U3 & 1.159 & 1.409 & 10.819 & 2.293 & 1.500 & 0.028 \\
		 TK2U4 & 0.954 & 1.418 & 13.319 & 2.041 & 2.000 & 0.070 \\
	\end{tabular}
	\caption{\label{tablea2} Stellar properties of the equilibrium models of sequence TK2U constructed by the \texttt{XNS} code.
	TK2U is a sequence of uniformly rotating neutron star with a fixed rest mass $M_{0} = 1.506$ and toroidal magnetization constant $K_{\mathrm{m}}=1.0$.}
\end{table}

\begin{table}[H]
	\centering
	\begin{tabular}{ccccccc}
		 Model & $\rho_\mathrm{c}$ & $M$ & $R$ & $\mathcal{B}_\mathrm{max}$ & $\Omega_\mathrm{c}$ & $T/|W|$\\
        & ($10^{-3}$) &  &   & ($10^{17}$ G)  & ($10^{-2}$)  &    \\
		\hline
  	 TK3U0 & 1.344 & 1.408 & 10.094 & 3.474 & 0.000 & 0.000 \\
		 TK3U1 & 1.329 & 1.409 & 10.189 & 3.450 & 0.500 & 0.003 \\
		 TK3U2 & 1.282 & 1.411 & 10.521 & 3.374 & 1.000 & 0.011 \\
		 TK3U3 & 1.189 & 1.414 & 11.280 & 3.217 & 1.500 & 0.028 \\
		 TK3U4 & 1.112 & 1.417 & 12.177 & 3.079 & 1.750 & 0.043 \\
	\end{tabular}
	\caption{\label{tablea3} Stellar properties of the equilibrium models of sequence TK3U constructed by the \texttt{XNS} code.
	TK3U is a sequence of uniformly rotating neutron star with a fixed rest mass $M_{0} = 1.506$ and toroidal magnetization constant $K_{\mathrm{m}}=1.5$.}
\end{table}

\begin{table}[H]
	\centering
	\begin{tabular}{ccccccc}
		 Model & $\rho_\mathrm{c}$ & $M$ & $R$ & $\mathcal{B}_\mathrm{max}$ & $\Omega_\mathrm{c}$ & $T/|W|$\\
        & ($10^{-3}$) &  &   & ($10^{17}$ G)  & ($10^{-2}$)  &    \\
		\hline
  	 TK4U0 & 1.362 & 1.414 & 10.554 & 4.289 & 0.000 & 0.000 \\
		 TK4U1 & 1.347 & 1.414 & 10.650 & 4.258 & 0.500 & 0.003 \\
		 TK4U2 & 1.299 & 1.416 & 11.076 & 4.155 & 1.000 & 0.011 \\
		 TK4U3 & 1.204 & 1.419 & 12.117 & 3.942 & 1.500 & 0.029 \\
		 TK4U4 & 1.119 & 1.423 & 13.718 & 3.739 & 1.750 & 0.046 \\
	\end{tabular}
	\caption{\label{tablea4} Stellar properties of the equilibrium models of sequence TK4U constructed by the \texttt{XNS} code.
	TK4U is a sequence of uniformly rotating neutron star with a fixed rest mass $M_{0} = 1.506$ and toroidal magnetization constant $K_{\mathrm{m}}=2.0$.}
\end{table}

\begin{table}[H]
	\centering
	\begin{tabular}{ccccccc}
		 Model & $\rho_\mathrm{c}$ & $M$ & $R$ & $\mathcal{B}_\mathrm{max}$ & $\Omega_\mathrm{c}$ & $T/|W|$\\
        & ($10^{-3}$) &  &   & ($10^{17}$ G)  & ($10^{-2}$)  &    \\
		\hline
  	 TK5U0 & 1.362 & 1.420 & 11.203 & 4.902 & 0.000 & 0.000 \\
		 TK5U1 & 1.347 & 1.421 & 11.345 & 4.862 & 0.500 & 0.003 \\
		 TK5U2 & 1.297 & 1.422 & 11.913 & 4.732 & 1.000 & 0.012 \\
		 TK5U3 & 1.194 & 1.426 & 13.612 & 4.448 & 1.500 & 0.032 \\
	\end{tabular}
	\caption{\label{tablea5} Stellar properties of the equilibrium models of sequence TK5U constructed by the \texttt{XNS} code.
	TK5U is a sequence of uniformly rotating neutron star with a fixed rest mass $M_{0} = 1.506$ and toroidal magnetization constant $K_{\mathrm{m}}=2.5$.}
\end{table}

\begin{table}[H]
	\centering
	\begin{tabular}{ccccccc}
		 Model & $\rho_\mathrm{c}$ & $M$ & $R$ & $\mathcal{B}_\mathrm{max}$ & $\Omega_\mathrm{c}$ & $T/|W|$\\
        & ($10^{-3}$) &  &   & ($10^{17}$ G)  & ($10^{-2}$)  &    \\
		\hline
		 TK1D0 & 1.290 & 1.401 & 9.638 & 1.285 & 0.000 & 0.000 \\
		 TK1D1 & 1.284 & 1.401 & 9.639 & 1.283 & 0.500 & 0.001 \\
		 TK1D2 & 1.264 & 1.402 & 9.734 & 1.275 & 1.000 & 0.004 \\
		 TK1D3 & 1.229 & 1.403 & 9.831 & 1.260 & 1.500 & 0.009 \\
		 TK1D4 & 1.176 & 1.404 & 10.023 & 1.239 & 2.000 & 0.017 \\
		 TK1D5 & 1.096 & 1.407 & 10.405 & 1.204 & 2.500 & 0.031 \\
		 TK1D6 & 0.974 & 1.411 & 10.979 & 1.148 & 3.000 & 0.052 \\ 
		 TK1D7 & 0.864 & 1.416 & 11.646 & 1.094 & 3.500 & 0.073 \\  
	\end{tabular}
	\caption{\label{tablea6} Stellar properties of the equilibrium models of sequence TK1D constructed by the \texttt{XNS} code.
	TK1D is a sequence of differentially rotating neutron star with a fixed rest mass $M_{0} = 1.506$ and toroidal magnetization constant $K_{\mathrm{m}}=0.5$.}
\end{table}

\begin{table}[H]
	\centering
	\begin{tabular}{ccccccc}
		 Model & $\rho_\mathrm{c}$ & $M$ & $R$ & $\mathcal{B}_\mathrm{max}$ & $\Omega_\mathrm{c}$ & $T/|W|$\\
        & ($10^{-3}$) &  &   & ($10^{17}$ G)  & ($10^{-2}$)  &    \\
		\hline
		 TK5D0 & 1.362 & 1.420 & 11.203 & 4.902 & 0.000 & 0.000 \\
		 TK5D1 & 1.354 & 1.421 & 11.250 & 4.885 & 0.500 & 0.001 \\
		 TK5D2 & 1.332 & 1.421 & 11.393 & 4.832 & 1.000 & 0.005 \\
		 TK5D3 & 1.292 & 1.423 & 11.725 & 4.735 & 1.500 & 0.012 \\
		 TK5D4 & 1.225 & 1.425 & 12.246 & 4.568 & 2.000 & 0.024 \\
		 TK5D5 & 1.103 & 1.429 & 13.617 & 4.240 & 2.500 & 0.048 \\
	\end{tabular}
	\caption{\label{tablea7} Stellar properties of the equilibrium models of sequence TK5D constructed by the \texttt{XNS} code.
	TK5D is a sequence of differentially rotating neutron star ($\Tilde{A}=1.0$) with a fixed rest mass $M_{0} = 1.506$ and toroidal magnetization constant $K_{\mathrm{m}}=2.5$.}
\end{table}


\bibliography{references}{}

\end{document}